\documentstyle[12pt,iopconf,epsfig]{article}

\begin{document}

\title{Neutrino-driven supernovae:  Boltzmann neutrino transport and the 
explosion mechanism}

\author{O. E. B. Messer\dag\ddag\,
A. Mezzacappa\dag\ddag\, 
S. W. Bruenn\S, and \\
M. W. Guidry\dag\ddag\
}

\affil{\footnotesize \dag\ Physics Division, Oak Ridge National
Laboratory,  Oak Ridge, TN 37831}
\affil{\footnotesize \ddag\ Department of Physics and Astronomy, University of Tennessee,
 Knoxville, TN 37996}
\affil{\footnotesize \S\ Department of Physics, Florida Atlantic University, Boca Raton, FL 33431}

\beginabstract
\begin{small}
Core-collapse supernovae are, despite their spectacular visual display, neutrino 
events.  Virtually all ($\sim$99\%) of the $10^{53}$ ergs of gravitational binding energy released in 
the formation of the nascent neutron star is carried away in the form of neutrinos 
and antineutrinos of  
all three flavors,
and these neutrinos are primarily responsible for powering the explosion.  This mechanism 
depends sensitively on the neutrino transport between the neutrinospheres and the 
shock.  In light of this, we
   have performed a comparison of multigroup Boltzmann 
neutrino transport ({\footnotesize MGBT})   
and  (Bruenn's) multigroup flux-limited 
diffusion ({\footnotesize MGFLD}) in post-core bounce            
environments.  
Our analysis concentrates on those quantities central to 
the postshock matter heating stemming from electron neutrino 
and antineutrino absorption, namely 
the neutrino luminosities, {\footnotesize RMS} energies, and 
mean inverse flux factors.  We show that 
{\footnotesize MGBT} yields mean inverse 
flux factors in the gain region that are $\sim$25\% larger and luminosities that are 
$\sim$10\% larger
than those computed by {\footnotesize MGFLD}. 
Differences in the mean inverse flux factors, luminosities, and {\footnotesize RMS} energies
  translate to 
heating rates that are up to 2 times larger for Boltzmann 
transport, with net cooling rates below the gain radius that 
are typically $\sim$0.8 times the {\footnotesize MGFLD} rates.  These differences  are
 greatest at earlier postbounce times for a given progenitor mass, 
and for a given postbounce time, 
 greater for greater progenitor mass.
The increased differences  with
increased progenitor mass suggest that the net heating enhancement from
{\footnotesize MGBT} is potentially robust and self-regulated.

\end{small}
\endabstract

\section{\bf Introduction}

Ascertaining the core collapse supernova mechanism is a long-standing
problem in astrophysics. The current paradigm begins with the collapse of a 
massive star's iron core and the generation of an outwardly propagating shock 
wave that results from core rebound. Because of nuclear dissociation 
and neutrino losses, 
the shock stagnates. This sets the stage for a shock reheating mechanism 
whereby neutrino energy deposition via electron neutrino 
and antineutrino  
absorption on nucleons behind the shock reenergizes it (Bethe \& Wilson 1985; 
Wilson 1985). 

The shock reheating phase is essential to the supernova's success, but it 
is precisely this phase that is difficult to simulate realistically. During
shock reheating, core electron neutrinos and 
antineutrinos are radiated from their respective neutrinospheres, 
and a small fraction of this radiated energy is absorbed in the exterior 
shocked mantle. The shock reheating depends sensitively on the electron neutrino 
and antineutrino luminosities,  
spectra (best characterized by the {\small RMS} energies),
 and angular distributions in the region 
behind the shock (e.g., see Burrows \& Goshy 1993, Janka \& M\"{u}ller 1996,
 Mezzacappa \etal 1998).
  
These, in turn, depend on the neutrino transport in the 
semitransparent region encompassing the neutrinospheres, necessitating a 
neutrino transport treatment that is able to transit accurately and 
seamlessly between neutrino-thick and neutrino-thin regions. 

Various neutrino transport approximations have been implemented in simulating
core collapse supernovae.
The most sophisticated approximation, which
naturally has been used in realistic one-dimensional simulations, is multigroup 
flux-limited diffusion ({\small MGFLD}; e.g., Bowers \& Wilson 1982, Bruenn 1985, 
Myra \etal 1987). 
{\small MGFLD} closes the neutrino radiation hydrodynamics hierarchy of equations 
at the level of the first moment (the neutrino flux) by imposing a relationship
between the flux and the gradient of the neutrino energy density (the zeroth
moment).  For example, 

\begin{equation}
F_{\nu}=-\frac{c\Lambda}{3}\frac{\partial U_{\nu}}{\partial r}+...,
\label{eq:mgfld}
\end{equation}

\begin{equation}
\Lambda = \frac{1}{1/\lambda + |\partial U_{\nu}/\partial r|/3U_{\nu}},
\label{eq:lambda}
\end{equation}
\smallskip

\noindent where $\lambda$ is the neutrino mean free path, and $U_{\nu}$ and $F_{\nu}$ 
are the neutrino energy density and flux (Bruenn 1985). 
[Other forms for the flux-limiter $\Lambda$ can be found in Bowers \& Wilson (1982),
Levermore \& Pomraning (1981),
and Myra \etal (1987).]
Whereas the limits $\lambda \rightarrow 0$
  and $\lambda \rightarrow \infty$  produce
the correct diffusion and free streaming fluxes, it is in the critical intermediate 
regime where the {\small MGFLD} approximation is of unknown accuracy.  
Unfortunately, the quantities central 
to the postshock neutrino heating, i.e., the neutrino {\small RMS} energies,
 luminosities, and mean inverse flux factors,
 are determined in this 
regime, and given the sensitivity of the neutrino heating to these quantities, it becomes 
necessary to consider more accurate transport schemes.  Moreover, in detailed one-dimensional 
simulations that have implemented elaborate {\small MGFLD} neutrino transport (e.g., see
 Bruenn 1993, Wilson \& Mayle 1993, and Swesty \& Lattimer 1994), explosions were not 
obtained unless the neutrino heating was boosted by additional phenomena, such as convection. 
This leaves us with at least two possibilities to consider:  (1) Failures to produce 
explosions in the absence of additional phenomena, such as convection, have resulted 
from neutrino transport approximation.  (2) Additional phenomena may be essential 
in obtaining explosions.  

\section{Initial Models, Codes, and Methodology}

We begin with 15 ${\rm M}_{\odot}$ and 25 ${\rm M}_{\odot}$ precollapse models
 S15s7b and S25s7b 
provided by Woosley (1995). The initial models were evolved through 
core collapse and bounce using one-dimensional Lagrangian hydrodynamics and 
{\small MGFLD} neutrino transport  
coupled to  the Lattimer-Swesty equation of
state (Lattimer \& Swesty 1991). The data at 106 ms 
 and
233 ms after bounce for S15s7b and 156 ms after bounce for S25s7b were thermally 
 and hydrodynamically frozen. 
Stationary-state neutrino distributions were computed for
these profiles using both {\small MGBT} and 
{\small MGFLD}. 

The {\small MGBT} simulations were performed using
{\small BOLTZTRAN}: a Newtonian gravity,  $O(v/c)$, three-flavor, Boltzmann neutrino transport
code developed for the supernova problem and used thus far for studies of stellar 
core collapse (Mezzacappa and Matzner 1989, Mezzacappa \& Bruenn 1993abc). The {\small
MGFLD} simulations were performed using {\small MGFLD-TRANS}:
a Newtonian gravity,  $O(v/c)$,
 three-flavor, {\small MGFLD} neutrino transport code, which
has been used for both core collapse and postbounce evolution
(e.g., Bruenn 1985, 1993).

The {\small MGBT} simulations used 110 nonuniform radial spatial 
zones  
and 12 neutrino energy 
zones spanning a range between 5 and 225 MeV. The 
{\small MGFLD} used the same spatial and energy grids.  
Simulations with 
20 energy zones spanning the same energy range were performed with 
{\small BOLTZTRAN}; no changes in the results presented here were seen.

For the {\small MGBT} simulations there is an added dimension:  neutrino direction
cosine.  Because {\small MGBT} computes the neutrino distributions
as a function of direction cosine and energy for each spatial zone, the isotropy
of the neutrino radiation field as
a function of radius and neutrino energy is computed from first principles.
This is one of the key features distinguishing {\small MGBT} and {\small MGFLD}.  Because
the isotropy of the neutrino radiation field is critical to the shock reheating/revival, four Gaussian
quadrature sets (2--, 4--, 6--, and 8--point) were implemented in the {\small MGBT} simulations to ensure
numerical convergence of the results.

\section{\bf Results}

For electron neutrino and antineutrino absorption on neutrons and protons, 
the neutrino heating rate (in MeV/nucleon) in the region between the neutrinospheres 
and the shock can be written as

\begin{equation}
\dot{\epsilon}=\frac{X_{n}}{\lambda_{0}^{a}}\frac{L_{\nu_{\rm e}}}{4\pi r^{2}}
                <E^{2}_{\nu_{\rm e}}><\frac{1}{\sf F}>
              +\frac{X_{p}}{\bar{\lambda}_{0}^{a}}\frac{L_{\bar{\nu}_{\rm e}}}{4\pi r^{2}}
                <E^{2}_{\bar{\nu}_{\rm e}}><\frac{1}{\bar{\sf F}}>
\label{eq:heatrate}
\end{equation} 
\smallskip

\noindent where: $\lambda_{0}^{a}=\bar{\lambda}_{0}^{a}={G_{F}^2}{\rho}{(g_{V}^2+3g_{A}^2)}/{\pi}{(hc)^4}{m_{B}}$;
$G_{F}/(\hbar c)^{3}=1.166\times10^{-5}$ GeV$^{-2}$ is the Fermi 
coupling constant; $\rho$ is the 
matter density; 
$g_{V}=1.0$, $g_{A}=1.23$; $m_{B}$ is the baryon mass; $X_{n,p}$ are the free neutron and proton mass fractions; 
and $L_{\nu_{\rm e},\bar{\nu}_{\rm e}}$, $<E^{2}_{\nu_{\rm e},\bar{\nu}_{\rm e}}>$, 
and ${\sf F},\bar{\sf F}$ are the electron neutrino and antineutrino luminosities, {\small RMS} 
energies, and mean inverse flux factors, defined by

\begin{equation}
L_{\nu_{\rm e}}=4\pi r^{2}\frac{2\pi c}
                               {(hc)^{3}}\int dE_{\nu_{\rm e}} d\mu_{\nu_{\rm e}} E^{3}_{\nu_{\rm e}} \mu_{\nu_{\rm e}} f, 
\label{eq:nuelumin}
\end{equation}

\begin{equation}
\langle E^{2}_{\nu_{\rm e}}\rangle = \frac{\int dE_{\nu_{\rm e}}d\mu_{\nu_{\rm e}} E^{5}_{\nu_{\rm e}} f}
                                          {\int dE_{\nu_{\rm e}}d\mu_{\nu_{\rm e}} E^{3}_{\nu_{\rm e}} f}, 
\label{eq:nuerms}
\end{equation}

\begin{equation}
\langle \frac{1}{\sf F}\rangle =\frac{\int dE_{\nu_{\rm e}} d\mu_{\nu_{\rm e}} E^{3}_{\nu_{\rm e}} f}
                                     {\int dE_{\nu_{\rm e}} d\mu_{\nu_{\rm e}} E^{3}_{\nu_{\rm e}} \mu_{\nu_{\rm e}} f}
=\frac{cU_{\nu_{\rm e}}}{F_{\nu_{\rm e}}}. 
\label{eq:nuefluxfac}
\end{equation}
\smallskip

\noindent In equations (\ref{eq:nuelumin})--(\ref{eq:nuefluxfac}), 
$f$ is the electron neutrino distribution function, which is a function
of the electron neutrino direction cosine, $\mu_{\nu_{\rm e}}$, and energy, $E_{\nu_{\rm e}}$.
In equation (\ref{eq:nuefluxfac}), $U_{\nu_{\rm e}}$ and $F_{\nu_{\rm e}}$
are the electron neutrino energy density and flux. Corresponding quantities 
can be defined for the electron antineutrinos. 
Success or failure to generate explosions via neutrino reheating must ultimately rest on the 
three quantities defined in equations (\ref{eq:nuelumin})--(\ref{eq:nuefluxfac}). 

\begin{figure}[tb]
\begin{center}
\vspace{-12pt}
\epsfig{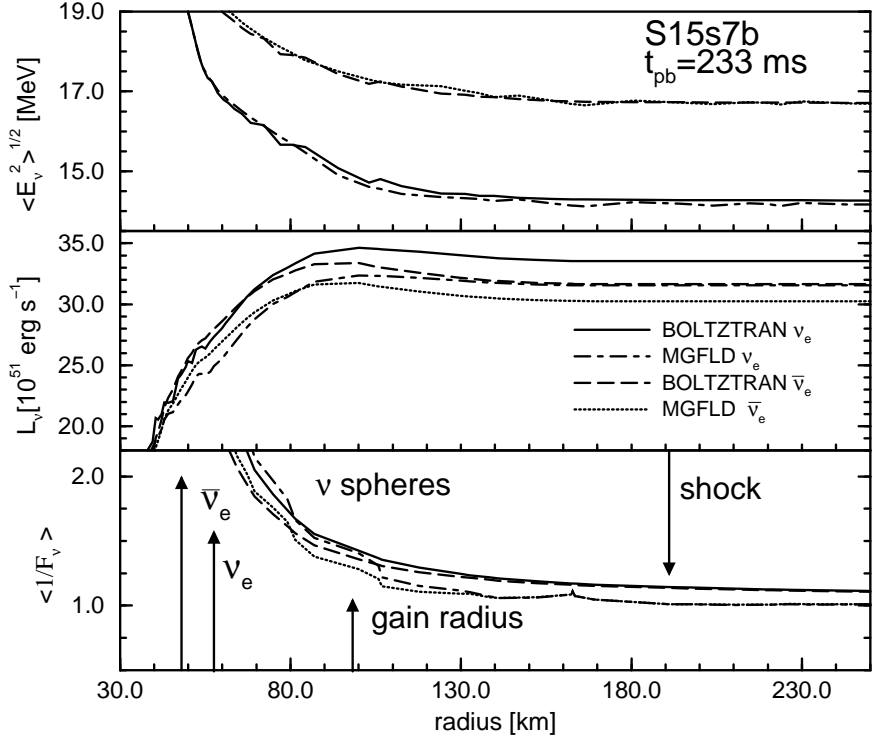}
\vspace{-12pt}
\begin{small}
\caption{RMS energies, luminosities, and mean inverse flux factors for model S15s7b 
at 233 ms after 
core bounce.}
\end{small} 
\label{fig:3pan}
\vspace{-28pt}
\end{center}
\end{figure}

In Figure 1, at 233 ms after bounce for model S15s7b, 
 we plot the electron neutrino and antineutrino {\small RMS} energies, luminosities, and mean 
inverse flux factors as a function of radius for our 
(8-point Gaussian quadrature) {\small MGBT} 
 and {\small MGFLD}
runs.
The  energy-averaged neutrinospheres 
(at 57 km and 48 km, for electron neutrinos and antineutrinos, respectively),
 and the location of the shock (at 191 km), are indicated by arrows. 
The gain radius (neutrino-energy integrated), located at 98 km, is also marked by an arrow.
For the electron neutrinos, the differences in {\small RMS} 
energies between {\small MGBT} and {\small MGFLD} are at most 
2\% throughout most of the region plotted, although  
{\small MGBT} consistently gives higher energies.  The 
differences between {\small MGBT} and {\small MGFLD} antineutrino {\small RMS} energies
are smaller, and neither transport scheme yields consistently higher values.  
It should be noted that we expect larger
differences when a fully hydrodynamic simulation is carried
out, with {\small MGBT} giving harder spectra (Mezzacappa and Bruenn 1993a,c;
 see also Burrows 1998). In
a static matter configuration,
differences that result from different treatments of the neutrino energy
shift measured by comoving observers do not occur.

Significant differences between {\small MGBT} and {\small MGFLD} are evident when comparing 
the neutrino and antineutrino luminosities and mean inverse flux factors.  
Both transport methods compute similar electron neutrino luminosities
until the neutrinospheres are approached from below.  The antineutrino 
luminosities for each transport method also coincide up to this point.  Just below the neutrinospheres, 
the {\small MGBT} luminosities diverge upward from the {\small MGFLD} luminosities, 
as {\small MGFLD} underestimates the neutrino flux, 
differing by 7\% (4\% for antineutrinos) at the neutrinospheres.  After a decline from this 
maximum difference, the fractional 
difference grows from approximately 
3\% at the base of the gain region to
a constant difference of 6\% beyond about 170 km.  Similar behavior is 
exhibited by the antineutrino luminosities, 
with the same fractional differences, 3\% and 6\%, obtained at the base 
of the gain region and near the shock, 
respectively.  

For the electron neutrinos, the fractional difference between $<1/{\sf F}>_{\rm \tiny MGFLD}$ and
$<1/{\sf F}>_{\rm \tiny MGBT}$
  is 2\%, 8\%, and 12\% at the neutrinosphere, gain radius, and shock, respectively.
Just above the shock, the difference converges to 10\%
 and is maintained to the edge of the core.
Focusing on the semitransparent region, 
 $<1/{\sf F}>_{\rm \tiny MGFLD}$ is greater  
below the gain radius until the gain radius is approached; i.e., the 
{\small MGFLD} neutrino radiation field is more isotropic 
than the {\small MGBT} radiation field in this region.  
At 80 km, as the gain radius is approached, {\small MGFLD}
computes a sharp decrease in $<1/{\sf F}>$.  This sharp decrease marks the radius at 
which the electron neutrino source is enclosed. 
The dip at 106 km and the sharp spike at 163 km in $<1/{\sf F}>_{\rm \tiny MGFLD}$ are
caused by local density perturbations.

For the electron antineutrinos, the same features are seen in $<1/{\sf F}>_{\rm \tiny MGFLD}$. 
  The fractional difference  
is 0\%, 11\%, and 11\% at the neutrinosphere, gain radius, and shock, respectively.
 The initial sharp decrease in 
$<1/{\sf F}>_{\rm \tiny MGFLD}$ occurs at a smaller radius;  
  the point at which the electron antineutrino source is enclosed 
is at a smaller radius.
These results are typical of all three time slices. 

Because each of the quantities plotted in Figure 1 is consistently greater for 
{\small MGBT}  (while this is not strictly true
for the antineutrino {\small RMS} energies in our stationary state comparisons, in a fully
dynamical simulation these energies will be consistently higher for {\small MGBT}
[Mezzacappa and Bruenn 1993a,c;
 see also Burrows 1998]),
and because the neutrino 
heating rate is proportional to each of them, 
 {\small MGBT} yields a significantly higher heating rate.  As an example, just above the
gain radius for model S15s7b, at $t_{\rm pb}=233$ ms and the net-heating peak,  
{\small MGBT} yields a heating rate from neutrino absorption that is $(102\%)^{2} \times 110\% \times 112\%$
 of the {\small MGFLD} rate. 

\begin{figure}[tb]
\begin{center}
\vspace{-12pt}
\epsfig{figure=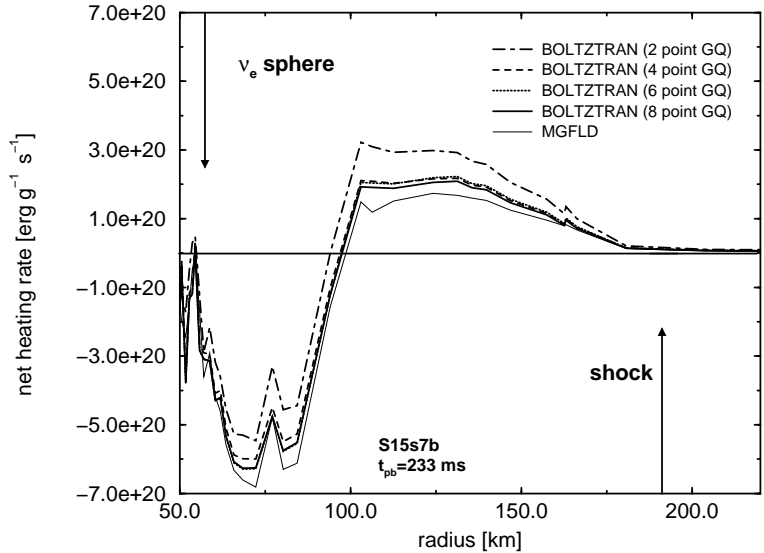,height=3.0in,clip=}
\vspace{-12pt}
\begin{small}
\caption{Net heating rates for model S15s7b at 233 ms after
core bounce.}
\end{small}
\label{fig:netheat}
\vspace{-28pt}
\end{center}
\end{figure}

In 
Figure 2, for   
{\small MGBT} and {\small MGFLD}, 
we plot the {\it net} neutrino heating rates as a function of radius 
for model S15s7b at $t_{\rm pb}=233$ ms. 
(As discussed in Section 2, the results from four Gaussian quadrature 
sets are plotted to demonstrate numerical
convergence.) 
These rates include the
contributions from both the electron neutrinos and antineutrinos, and were computed using 
the following formulae: 

\begin{equation}
(d\epsilon /dt)_{i}=c\int E_{\nu}^{3}dE_{\nu}
[\psi^{0}_{i}/\lambda^{(a)}_{i}-j_{i}(1-\psi^{0}_{i})]/\rho (hc)^{3} 
\end{equation}

\noindent where $\epsilon$ is the internal energy per gram; 
$E_{\nu}$, $\psi^{0}_{i}$, $\lambda^{(a)}_{i}$, and $j_{i}$ 
are the electron neutrino or antineutrino energy, zeroth angular 
moment, absorption mean free path, 
and emissivity, respectively; 
$i=1$ corresponds to electron neutrinos, and 
$i=2$ corresponds to electron antineutrinos. 
Only the contributions from neutrino emission and absorption
were included. 
The {\small MGBT} simulation yields a net heating rate just above the gain radius
 that is 
$\sim$1.3 times the {\small MGFLD} rate, and a net cooling rate
below the gain radius that is consistently $\sim$0.8 times the {\small MGFLD} rate. 
The differences in net heating rate are even greater for the $t_{\rm pb}=106$ ms time slice in
our 15 M$_{\odot}$ model  and for the $t_{\rm pb}=156$ ms time slice in 
our 25 M$_{\odot}$  model (cf. Table 1 and Messer \etal 1998).

\section{\bf Summary, Discussion, and Conclusions}

Comparing three-flavor {\small MGBT} and three-flavor 
{\small MGFLD} in postbounce supernova environments, we find that 
 {\small MGBT} leads to a significant increase/decrease in the 
 {\it net} heating/cooling rate, particularly  just above/below 
 the gain radius. The {\small MGBT} net heating  rate can be 
as much as 2 times the {\small MGFLD} net heating rate above the gain radius, with
net cooling rates that are typically 0.8 times the {\small MGFLD} rate below. 
 These differences 
 stem primarily from differences in the neutrino luminosities and mean inverse flux 
factors;  the heating rate 
is linearly proportional to both these quantities, and differences in both add to 
produce a significant difference in the net heating rate. 

We also observe that the differences in the net heating rate are greatest at 
earlier postbounce times for a given progenitor mass, 
 and at a given postbounce time, greater for greater progenitor mass. 
This is illustrated in Table 1.  
  The enhancement in heating with increased progenitor mass
suggests that the net heating enhancement from {\small MGBT} is
potentially robust and self-regulated.

In closing, our results are promising, and their ramifications 
for core collapse supernovae,  and in particular, for the postbounce 
neutrino-heating, shock-revival mechanism, await one- and two-dimensional 
dynamical simulations with {\small MGBT} coupled 
to the core hydrodynamics. One-dimensional simulations are currently 
underway, and we plan to report on them soon.

\begin{table}[t]
\vspace{-24pt}
\begin{center}
\footnotesize\rm
  \caption{Maximum Net Heating/Cooling Rates \label{t1}}
      \begin{tabular}{lccc}
  \topline
            {Progenitor Mass [${\rm M}_{\odot}$]} &
            {$t_{pb}$ [ms]} &
            {Maximum Net Heating Ratio} &
            {Maximum Net Cooling Ratio}
       \\
      \midline
            15&
            106&
            2.0&
            0.8
        \\  
             &
            233&
            1.3&
            0.8
        \\      
            25 &
            156&
            2.0&
            0.8
      \\
   \bottomline
\end{tabular}
\end{center}
\vspace{-12pt}
\end{table}

\section{Acknowledgements}
\begin{footnotesize}
BM
was supported at the University of Tennessee under NASA grant 
NRA-96-04-GSFC-073. AM and MWG were supported at the Oak Ridge National Laboratory,
which is managed by Lockheed Martin Energy Research Corporation
under DOE contract DE-AC05-96OR22464, and at the University of
Tennessee, under DOE contract DE-FG05-93ER40770.
SWB was supported at Florida Atlantic University 
under NASA grant NRA-96-04-GSFC-073 and NSF grant AST-9618423. 
The simulations presented in this paper were carried out on
the Cray C90 at the National Energy Research Supercomputer
Center, the Cray Y/MP at the North Carolina Supercomputer 
Center, and the Cray Y/MP and Silicon Graphics Power
Challenge at the Florida Supercomputer Center. AM and SWB 
gratefully acknowledge the hospitality of the Institute
for Theoretical Physics, Santa Barbara, 
which is supported in part by
the National Science Foundation under grant number
PHY94-07194. 
\end{footnotesize}

\end{document}